\documentclass[amsmath,amssymb,nofootinbib,twocolumn,aps,prl]{revtex4-1}
\usepackage{xcolor}
\usepackage{graphicx}
\usepackage{amsmath}
\begin{document}
\title{Enhanced Transport of Two Spheres in Viscous Fluid}
\author{Julian Lee}
\email{jul@ssu.ac.kr}
\affiliation{Department of Bioinformatics and Life Science, Soongsil University, Seoul 06978, Korea}
\author{Sean L. Seyler}
\affiliation{Department of Physics, Arizona State University, Tempe, Arizona 85287, USA}
\author{Steve Press\'e}
\email{spresse@asu.edu}
\affiliation{Department of Physics, Arizona State University, Tempe, Arizona 85287, USA}
\date{\today}
\begin{abstract}
	We obtain a numerical solution for the synchronous motion of two spheres moving in viscous fluid. We find that for a given amount of work performed, the final distance travelled by each sphere is increased by the presence of the other sphere. The result suggests that the transport efficiency of molecular motor cargo {\it in vivo} may be improved due to an effective hydrodynamic interaction with neighboring cargos moving along the same direction.   
\end{abstract}
\maketitle
Subcellular transport processes have been the subject of continuing interest largely because of the high efficiency of
molecular-scale motors responsible for such transport, as compared to the efficiency of their macroscopic counterparts~\cite{eff,eff2,eff3,eff4,keff,motor1,motor2,kol2,neuman,hyeon1,motor3,motor4,motor5,motor6,hyeon2}.
An important factor affecting such a transport process is the hydrodynamic interactions~\cite{add1,add2,add3,add4,add5,khei,hyd1,hyd2,hyd3,hyd4,hyd5,hyd6,1sp,2sp,jo1,jo2,fh1,fh2,fh3,fh4} of the molecular motor and cargo with the surrounding viscous fluid~\cite{kin5}. 
First, it has been shown that the Stokes drag experienced by a spherical object embedded in a fluid is reduced in the presence of another sphere in the vicinity~\cite{jf}. 
The reduction of the drag can be understood intuitively as coming from the indirect transfer of the momentum between the embedded objects, mediated by the fluid. Secondly, there is a separate effect of hydrodynamic memory due to the unsteady flow~\cite{bous,basset,oseen,khei,hyd1,hyd2,hyd3,hyd4,hyd5,hyd6,1sp,2sp}, where the momentum transferred from an embedded object is transiently stored in the fluid, and then transferred back to the object at a later point~\cite{1sp}. Such an effect may facilitate the transport even for a single spherical object~\cite{onesph}.

These considerations motivate us to ask whether fluid flow generated by the motion of a moving object in viscous fluid indeed facilitates the transport of a neighboring cargo. 
To address this question here, we examine the displacement of two neighboring spheres under the same constant external force of finite duration by numerically integrating the equation of motion~\cite{2sp} (Eq.(\ref{bbo22})). 

We will show that for a given distance traveled, the required input work for two neighboring spheres is less than that for spheres separated by large distances. We also find that two neighboring spheres are transported faster than separated spheres, supporting the idea that indirect, fluid-induced, inter-vesicle interaction may be utilized in active subcellular transport for improved efficiency. 

\begin{figure}
\includegraphics[width=0.8\textwidth]{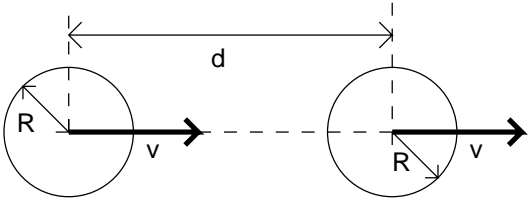}
\caption{The spheres moving along the center-to-center axis. The directions of the velocity are shown by the bold arrows which are also the directions of the external forces. The center-to-center distance and the radius are $d$ and $R$, respectively.}
\label{sch1} 
\end{figure}

{\it Equation for two spheres}--
The equation of motion for two spheres moving in viscous fluid has previously been derived by generalizing the equation for one sphere in a fluid~\cite{2sp}. Briefly, fluid flow is computed in the presence of a sphere of radius $R$ with no-slip boundary condition by using the unsteady Stokes equation where the convective term in the Navier-Stokes equation is neglected but the time-derivative is kept~\cite{LL}.  Another sphere is introduced at a center-to-center distance $d$, then the flow is perturbed in order to satisfy the no-slip boundary at the second sphere, and perturbed again in order to satisfy the boundary condition at the first sphere. This process is iteratively repeated. The series is truncated so that the resulting error is of order 
$\epsilon^4$ where $\epsilon \equiv R/d$~\cite{2sp}.  The resulting expression for the force has a remarkable symmetry with respect to the exchange of the spheres' positions. In particular, when their radii are the same and they move with the same velocity ${\bf v}(t)$, the force exerted on each of the spheres is the same. In the special case of one-dimensional motion, along a line connecting their centers, with the same velocity component 
$v(t)$~(Fig.~\ref{sch1}), the equation reduces to~\cite{2sp} 
\begin{eqnarray}
	F_{\rm fluid} &=&  - \frac{ 6 \pi \eta R }{1+\frac{3}{2}\epsilon - \epsilon^3} v
 -  \frac{ 2 \pi R^3 \rho_f}{3(1+3 \epsilon^3)} \dot v \nonumber\\
&&- 6 \pi \eta R  \int^t_{-\infty} dt' \dot v(t')h_\epsilon(t-t'), \label{bbo21}
\end{eqnarray}
where $\rho_f$ and $\eta$ are the density and the dynamic viscosity of the fluid, respectively. The functional form of the memory kernel $h_\epsilon(t)$ is formally given as a Laplace transform~\cite{2sp}; see SI for details. 
A similar expression can be obtained for the spheres whose motion is in the direction perpendicular to the lines joining their centers (see SI). Because the force exerted on each sphere by the fluid is the same, the motion of both spheres can be kept synchronous by applying the same external force $F(t)$ under the same initial velocities. The equation of motion for each sphere can be written as 
\begin{eqnarray}
&&\left( \rho_s + \frac{\rho_f }{2(1+3 \epsilon^3)}  \right) \dot v =  - \frac{9 \eta}{2 R^2}  \frac{  v}{1+\frac{3}{2}\epsilon - \epsilon^3} \nonumber\\
&&- \frac{9 \eta}{2 R^2}    \int^t_{-\infty} dt' \dot v(t')h_\epsilon(t-t') + \frac{3}{4 \pi R^3} F(t), \label{bbo22}
\end{eqnarray}
where $\rho_s$ is the density of the sphere. By comparing with the exact result~\cite{jf} for the special case of the motion with constant velocity, it was argued that Eq.~(\ref{bbo22}) is a reasonable approximation for $\epsilon \lesssim 0.25$
~\cite{2sp}.
The functional form of $h_\epsilon(t)$ is such that $h_0(t)= R\sqrt{\frac{\rho_f}{\pi \eta t}}$ (see SI), so that for $\epsilon=0$, we obtain the familiar Basset-Boussinesq-Oseen equation~\cite{bous,basset,oseen,khei}
for an accelerating sphere in a fluid
\begin{eqnarray}
	&&(\rho_s + \frac{\rho_f }{2}) \dot v =  - \frac{9 \eta}{2 R^2} v \nonumber\\
	&&- \frac{9 }{2 R}  \sqrt{\frac{\eta \rho_f}{\pi}}\int^t_{-\infty} dt' \frac{\dot v(t')}{\sqrt{t-t'}} + \frac{3}{4 \pi R^3} F(t), \label{bbo}
\end{eqnarray}
where the second term on the right hand side captures the effect of the hydrodynamic memory~\cite{bous,basset,oseen,khei}.   

For convenience, we may rewrite the equation for synchronous movement of two spheres in terms of dimensionless quantities, defined as 
\begin{eqnarray}
	\theta &\equiv& t/\tau_B, \quad  u(\theta) \equiv 6\pi \eta R v(t)/ F_{\rm max},\nonumber\\
	f(\theta) &\equiv& F(t)/F_{\rm max},
\end{eqnarray}
where $F_{\rm max}$ is the maximum value of $F(t)$ and $\tau_B \equiv (2 \rho_s + \rho_f)R^2/9  \eta$ is the Brownian relaxation time~\cite{pad}. Then, Eq.~(\ref{bbo22}) is rewritten as
\begin{eqnarray}
	&&	\frac{2  \rho_s (1+3 \epsilon^3)+ \rho_f}{(1+3 \epsilon^3)(2  \rho_s +  \rho_f)} \frac{d u}{d \theta} = - \frac{u}{1+\frac{3}{2}\epsilon - \epsilon^3} \nonumber\\ &-&  \int^\theta_{-\infty} d \theta' \frac{d u}{d \theta'} h_\epsilon(\tau_B(\theta-\theta')) 
	+ f(\theta). \label{bbo3}
\end{eqnarray}
The velocity $u(\theta)$ can be obtained by numerical integration of Eq.~(\ref{bbo3}), by discretizing the time $\theta$. The integral can then be performed by a simple trapezoidal rule, which gives a reasonably accurate result when the discretization is performed with step $\Delta \theta=0.001$, as can be checked for the case of $\epsilon=0$ where exact solutions are available for certain special cases (see SI). Once $u(\theta)$ is obtained, it is then straightforward to obtain the non-dimensionalized quantities $x$ and $w$ corresponding to position and work, respectively, by additional integration: 
\begin{eqnarray}
	x(\theta) &\equiv& \int_0^\theta u(\theta') d\theta' + x(0) \equiv  \frac{6\pi \eta R}{F_{\rm max} \tau_B} X(t),\nonumber\\
	w(\theta) &\equiv& \int_0^\theta f(\theta') u(\theta') d\theta',
\end{eqnarray}
where $X(t)$ is the position of the sphere.
In order to quantify a notion of efficiency for transport, we define the dimensionless  effective transport drive force $f_{\rm drive}$~\cite{1sp},
\begin{equation}
	f_{\rm drive} \equiv \frac{w(\theta)}{\Delta x(\theta)}
\end{equation}
where $\Delta x(\theta) \equiv x(\theta)-x(0)$ is the displacement. This quantity is the dimensionless version of the specific energy consumption, often used for the measure of transport efficiency in the transportation industry~\cite{tr1,tr2,tr3}. The smaller value of $f_{\rm drive}(\theta)$ implies less amount of external work required for a given displacement. We also define the dimensionless effective friction~\cite{1sp} 
\begin{equation}
	z(\theta) \equiv \frac{f_{\rm drive}(\theta)}{{\bar u}(\theta)} = \frac{w(\theta)\Delta \theta}{\Delta x(\theta)^2}.
\end{equation}

\begin{figure}
\includegraphics[width=0.4\textwidth]{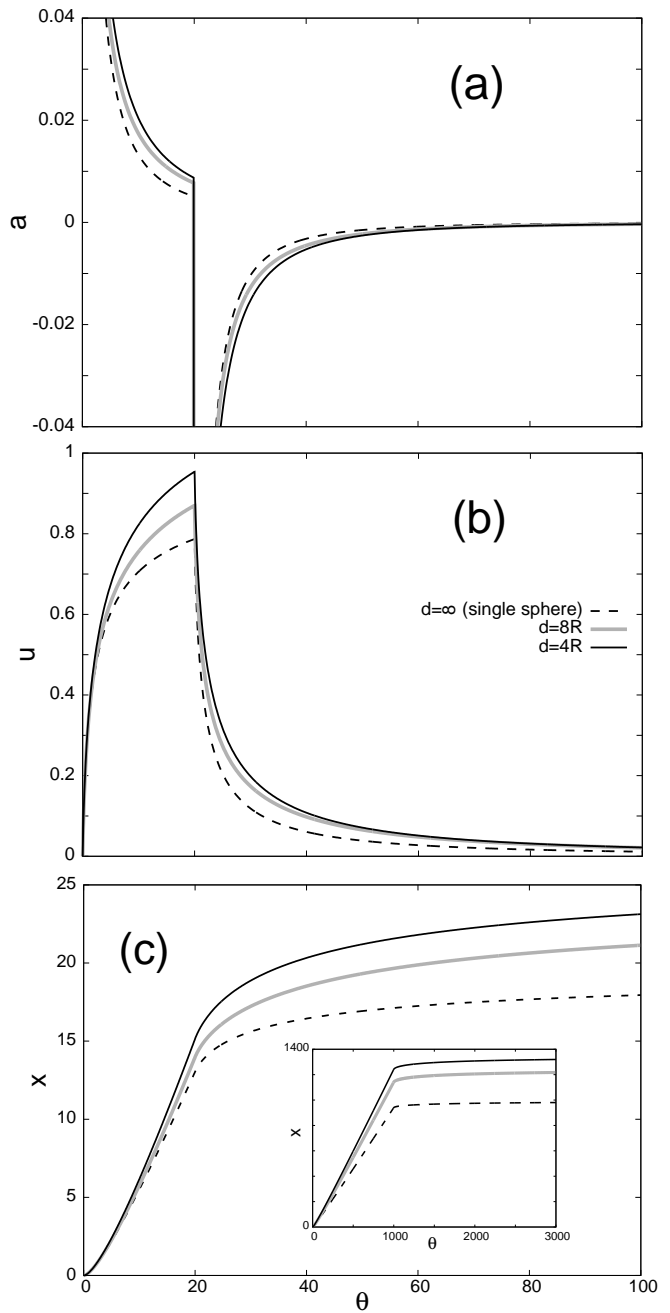}
	\caption{(a) The non-dimensional acceleration $a(\theta)$, (b) velocity $u(\theta)$, and (c) position $x(\theta)$, are compared for $d=\infty$ (single sphere), $d=8R$, and $d=4R$, for $T_{\rm pulse}=20 \tau_B$ and  $\rho_s=\rho_f$. The position for $T_{\rm pulse}=1000 \tau_B$ are shown in the inset of (c).}
\label{avx} 
\end{figure}

{\it Improved transport of two neighboring spheres}--
We now consider a simple protocol where a constant external force is applied over a finite duration $T_{\rm pulse}$, starting from $t=0$. Note that by definition, the maximum value of the normalized force $f(\theta)$ is unity. Therefore, $f(\theta)=1$ for $0 \le \theta \le \theta_{\rm pulse}$ 
and zero otherwise, where $\theta_{\rm pulse} \equiv T_{\rm pulse}/\tau_B$. We also take $x(0)=u(0)=0$. 
Therefore, the input work is $w(\theta)=f\times x(\theta)=x(\theta)$ for $\theta< \theta_{\rm pulse}$ and $w=x(\theta_{\rm pulse})$ otherwise. As such, 
\begin{equation}
	f_{\rm drive}(\theta) \equiv \frac{w(\theta)}{x (\theta)} 
	= \left\{
        \begin{array}{l}
		1,\ (\theta<\theta_{\rm pulse})\\       
		x(\theta_{\rm pulse})/x(\theta),\ (\theta \geq \theta_{\rm pulse}).
        \end{array}
        \right. \label{work}
\end{equation}
The results of the numerical computation for $\theta_{\rm pulse} = 20$ and $\rho_s=\rho_f$ are shown in Fig.~\ref{avx}, where the values of $a \equiv \frac{d u}{d \theta}$, $u$ and $x$ are compared for  $d=\infty$, $d=8R$, and $d=4R$. The case of $d=\infty$ corresponds to the motion of a single sphere. 
We find that for a {\it given strength} of the external force, the magnitudes of both acceleration and deceleration for two neighboring spheres are larger than  those for the spheres separated with a larger distance, and the overall effect is such that $v(t)$ for neighboring spheres is larger for all values of $t$. 

\begin{figure}
\includegraphics[width=0.4\textwidth]{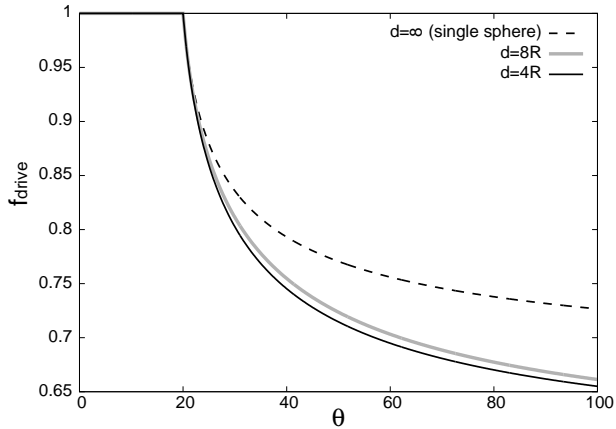}
	\caption{The effective transport drive force $f_{\rm drive}(\theta) \equiv w(\theta)/x(\theta)$ is compared for $d=\infty$ (single sphere), $d=8R$, and $d=4R$, for $T_{\rm pulse}=20 \tau_B$ and $\rho_s=\rho_f$.}
\label{eff} 
\end{figure}

\begin{figure}
\includegraphics[width=0.4\textwidth]{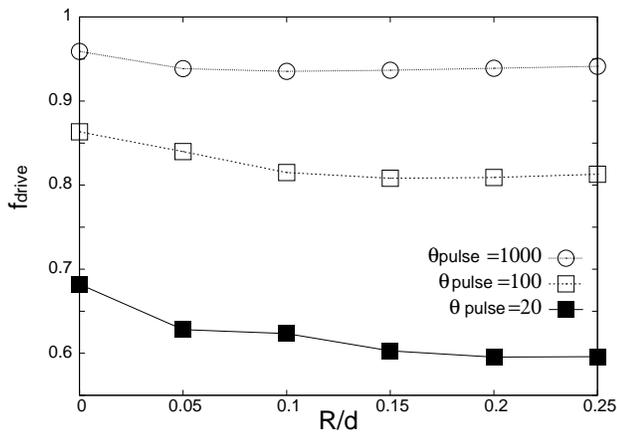}
	\caption{The final values of the effective transport drive force, $f_{\rm drive}(\infty) \equiv w(\infty)/x(\infty)$, are shown for several values of $R/d$ and $\theta_{\rm pulse} \equiv T_{\rm pulse}/\tau_B$, with $\rho_s=\rho_f$. The case of $R/d=0$ corresponds to that of a single sphere.} 
\label{egr} 
\end{figure}

More importantly, $x(\infty)$ increases by a large amount when the inter-sphere distance $d$ decreases, whereas $x(\theta_{\rm pulse})$ is only weakly dependent on distance. Therefore, from the second line of Eq. (\ref{work}), we find that $f_{\rm drive}(\infty)$ for neighboring spheres is less than that of a separate sphere. That is, the neighboring spheres travel farther compared to the separate spheres, for a given amount of input work. The graphs of $f_{\rm drive}(\theta)$ are shown for several values of $d$ in Fig.~\ref{eff}, where we see that in fact $f_{\rm drive}(\theta)$ is an increasing function $d$ for all values of $\theta$.
The values of $f_{\rm drive}(\infty)$ are also plotted in Fig.~\ref{egr} for several values of $R/d$ and $\theta_{\rm pulse}$, where 
approximate values of $x(\infty)$ are obtained from 
the values of $x(\theta)$ in those approximately flat regions of the $x$ plots (Fig.~\ref{avx}) with $u(\theta) \leq 0.01$. 
The trend of the reduced effective drive force for neighboring spheres is evident. 
In particular, we see that for $\theta_{\rm pulse}=20$, the value of $f_{\rm drive}$ reduces from 0.682 at $R=\infty$ to 0.596 at $d=4R$, resulting in about $\sim 13$ \% reduction in the required work for a given displacement. The reduction of $f_{\rm drive}$ is smaller for larger values of  $\theta_{\rm pulse}$.
This is because the motion of a sphere is diffusive at the time scale of $t \gg \tau_B$~\cite{khei}, with only tiny effects coming from hydrodynamic memory. This can be seen from the graphs of $x(\theta)$ for $\theta_{\rm pulse}=1000$, shown in the inset of Fig.~\ref{avx} (c), where the slope is approximately proportional to the instantaneous applied force, showing the typical behavior of an overdamped particle. In particular, since the motion of the sphere almost stops after the force pulse, we have $x(\theta_{\rm pulse}) \simeq x(\infty)$, leading to  $f_{\rm drive}(\infty) = x(\theta_{\rm pulse})/x(\infty) \simeq 1$ regardless of the inter-sphere distance, implying negligible reduction of the effective transport drive force.

\begin{figure}
\includegraphics[width=0.4\textwidth]{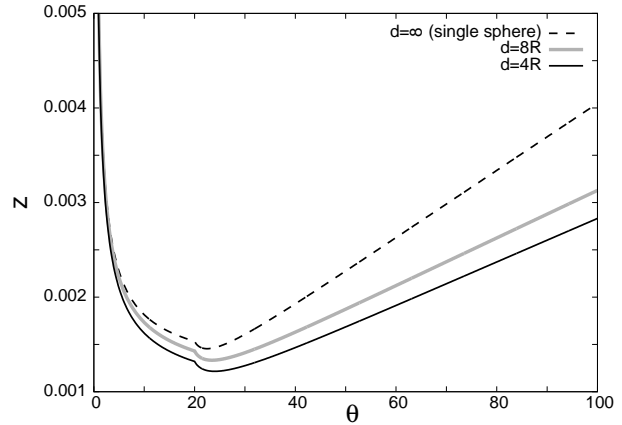}
	\caption{The graphs of the effective transport friction $z(\theta)$ are compared for $d=\infty$ (single sphere), $d=8R$, and $d=4R$, for $T_{\rm pulse}=20 \tau_B$ and $\rho_s=\rho_f$.}
\label{fr} 
\end{figure}

We also plot the graph of $z(\theta)$ in Fig.~\ref{fr} for $d=\infty$ (single sphere), $d=8R$, and $d=4R$, showing that for given displacements, two neighboring spheres not only requires less work but also results in faster transport as compared to separate spheres. The value of $z(\theta)$ here diverges as $\theta \to \infty$ because $\lim_{\theta \to \infty} \bar u(\theta) = 0$, but it will be maintained at finite values when a periodic force is applied so that a nonequilibrium steady state is reached~\cite{1sp}.


{\it Order of magnitude estimates with biological parameters}--
To see whether the reduced effective transport drive force and friction due to the indirect interaction between two spheres is relevant for subcellular transport processes, we perform an order of magnitude estimate using biological parameters. More specifically, we consider an example of cargo transport by a kinesin motor, where the constant force pulse can be considered as an  extremely simplified model of the force exerted by the kinesin motor and its cargo. The force duration may be taken as $T_{\rm pulse} = 10\ {\rm \mu s}$, the time scale during which the stepping motion occurs~\cite{kin3,kin4,kin5}. As discussed previously, the hydrodynamic memory plays a role only if $T_{\rm pulse}$ is not too much larger than $\tau_B \equiv (2 \rho_s + \rho_f)R^2/9  \eta$. For fixed values of $T_{\rm pulse}, \rho_s, \rho_f$, and $\eta$, this tells us that the size of the cargo should be sufficiently large 
in order for hydrodynamic memory to play a significant role. For example, we previously found that $f_{\rm drive}(\infty)$ 
for two spheres with $d = 4R$ was 13 \% less than for those separated by an infinite distance. Using the values $\eta = 2 \times 10^{-3}\  {\rm kg/(m\cdot s)}$~\cite{cy1,cy2,cy3,cy4} (BNID150903)\footnote{The ID number of BioNumbers Database~\cite{BN}.} and  $\rho_s=\rho_f = 10^3\ {\rm kg}/{\rm m}^3$~\cite{den1,den2} (BNID113851) for the cytoplasm, we get
\begin{equation}
T_{\rm pulse}=10\ {\mu \rm s} = 20 \tau_B = 20 \times \frac{ 3 \times  10^3\ {\rm kg}/{\rm m}^3 \times R^2}{ 9 \times  2 \times 10^{-3}\  {\rm kg/(m\cdot s)} }, 
\end{equation}
leading to
\begin{equation}
	R \simeq  2\ \mu {\rm m} \sim O(1 \mu {\rm m}),
\end{equation}
about the size of a large size vesicle such as an organelle~\cite{organ1,organ2,organ3}.
Considering the fact that organelles of size of order $1\ \mu \rm m$ are often transported by molecular motors~\cite{organ1,organ2,organ3}, the reduction of effective drive force and friction   driven by hydrodynamic effects warrants further investigation within living environments. 

{\it Discussion}--
In this work, we presented a numerical solution of two spheres moving in synchrony in a viscous fluid with the same force applied to each sphere. 
We found that for a given displacement for each sphere, the required work is less for two neighboring spheres than for spheres separated by a large distance, and the former is transported faster than the latter. 
In reality, the asynchrony in cargo transport may somewhat reduce the effect proposed here. Study of such a generalized case is straightforward, albeit technically more involved. 

Our results support the idea that the efficiency of subcellular transport may be improved by hydrodynamic interaction between the neighboring cargos. Taking thermal fluctuations explicitly into account, the transport efficiency is quantified by ~\cite{hyeon2,transp1}
\begin{equation}
	q \equiv \frac{Q(t) \langle \delta X(t)^2 \rangle}{\langle X(t) \rangle^2}\label{tf}
\end{equation}
where $Q(t)$ is the energy consumption up to time $t$, $\langle X(t) \rangle$ is the average displacement, and $\langle \delta X(t)^2 \rangle$ the transport precision. It has been shown that $q \ge 2 k_B T$, and this fundamental bound is called the thermodynamic uncertainty principle~\cite{un1,un2,un3,un4,un5,un6}. Within this bound, the molecular motor that performs transport with minimal energy expenditure and with highest precision is the most efficient one by definition. Since we considered the solution to a deterministic equation, the displacement we obtained  is expected to be the thermally averaged displacement. Since the input work is proportional to $Q(t)$ for a given value of thermal efficiency, and since our results tell us that $\langle X \rangle$ for two neighboring spheres is larger than that for separate spheres, Eq.(\ref{tf}) tells us that two neighboring spheres have higher transport efficiency due to hydrodynamic interactions, if $\langle \delta X(t)^2 \rangle$ are the same for both cases.
 Full analysis of the transport efficiency, taking into account transport precision in the presence of the thermal fluctuations, would require more sophisticated formalism such as the fluctuating hydrodynamics~\cite{fh1,fh2,fh3,fh4}.

\section{Acknowledgement} 
JL was supported by the National Research Foundation of Korea, funded by the Ministry of Education (NRF-2017R1D1A1B03031344). SS and SP were supported by ARO grant W911NF-17-1-0162 
on ``Multi-Dimensional and Dissipative Dynamical Systems: Maximum Entropy as a Principle for Modeling Dynamics and Emergent Phenomena in Complex Systems".

\bibliography{spheres}
\pagebreak
\widetext
\appendix
\section{Memory kernel for the two sphere equation (Eq.~(\ref{bbo22}))}\label{gbbo}
The memory kernel in Eq.~(\ref{bbo3}) is given by~\cite{2sp}
\begin{equation}
h_\epsilon(t) = \frac{1}{\pi}\int_0^\infty \hat h_\epsilon(s) e^{-s t/\tau_\nu} ds \label{int1}
\end{equation}
where
\begin{equation}
\hat h_\epsilon(s)={\rm Im}\left[\frac{A(s)}{B_\epsilon(s)}\right] 
\end{equation}
with 
 \begin{eqnarray}
A(s) &=&  \left(1 -  i \sqrt{s} - \frac{s}{9}\right)^2\nonumber\\
B_\epsilon(s) &=& s^2 \left(\frac{1}{9}+\frac{\epsilon^3}{3}\right) + i \left(1 + 2 \epsilon^3 \right) s^{3/2}       - \left(1+ 5  \epsilon^3 \right) s - 6 i \sqrt{s} \epsilon^3  + 3 \epsilon^3 \nonumber\\
&&+3 \epsilon^2 \exp\left(-i \sqrt{s}\left(1-\epsilon^{-1}\right)\right)\left[ 
     \frac{s^2}{90}  + \frac{i s^{3/2}}{6} - \frac{s}{2} - i \sqrt{s} + 1 \right]\left(i \sqrt{s} -\epsilon\right)
\end{eqnarray}
and $\tau_\nu \equiv \rho_f R^2/\eta$.
We note that for $\epsilon \to 0$, contribution from the term with the factor $\exp\left(-i \sqrt{s}\left(1-\epsilon^{-1}\right)\right)$ vanishes in the integral 
Eq.~(\ref{int1}), and we have
\begin{eqnarray}
h_\epsilon(t) &=& \frac{1}{\pi} \int_0^\infty {\rm Im}\left[\frac{(1 -  i \sqrt{s} - \frac{s}{9})^2}{ \frac{s^2}{9} + i  s^{3/2}       -  s}\right] e^{-s t/\tau_\nu} ds \nonumber\\
&=&  \frac{1}{\pi} \int_0^\infty \frac{1}{ \sqrt{s}} e^{-s t/\tau_\nu} ds = \sqrt{\frac{\tau_\nu}{\pi t}},
\end{eqnarray}
recovering the memory kernel for one sphere~\cite{bous,basset,oseen}.   

\section{The equation for spheres moving perpendicular to their line of centers }\label{perp}
\begin{figure}
\includegraphics[width=0.8\textwidth]{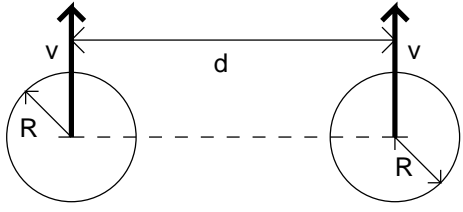}
\caption{Spheres moving perpendicular to the line connecting center-to-center. The directions of the velocities are shown by the bold arrows, which are also the directions of the external forces. The center-to-center distance and the radius are $d$ and $R$, respectively.}
\label{sch2} 
\end{figure}

\begin{figure}
\includegraphics[width=0.8\textwidth]{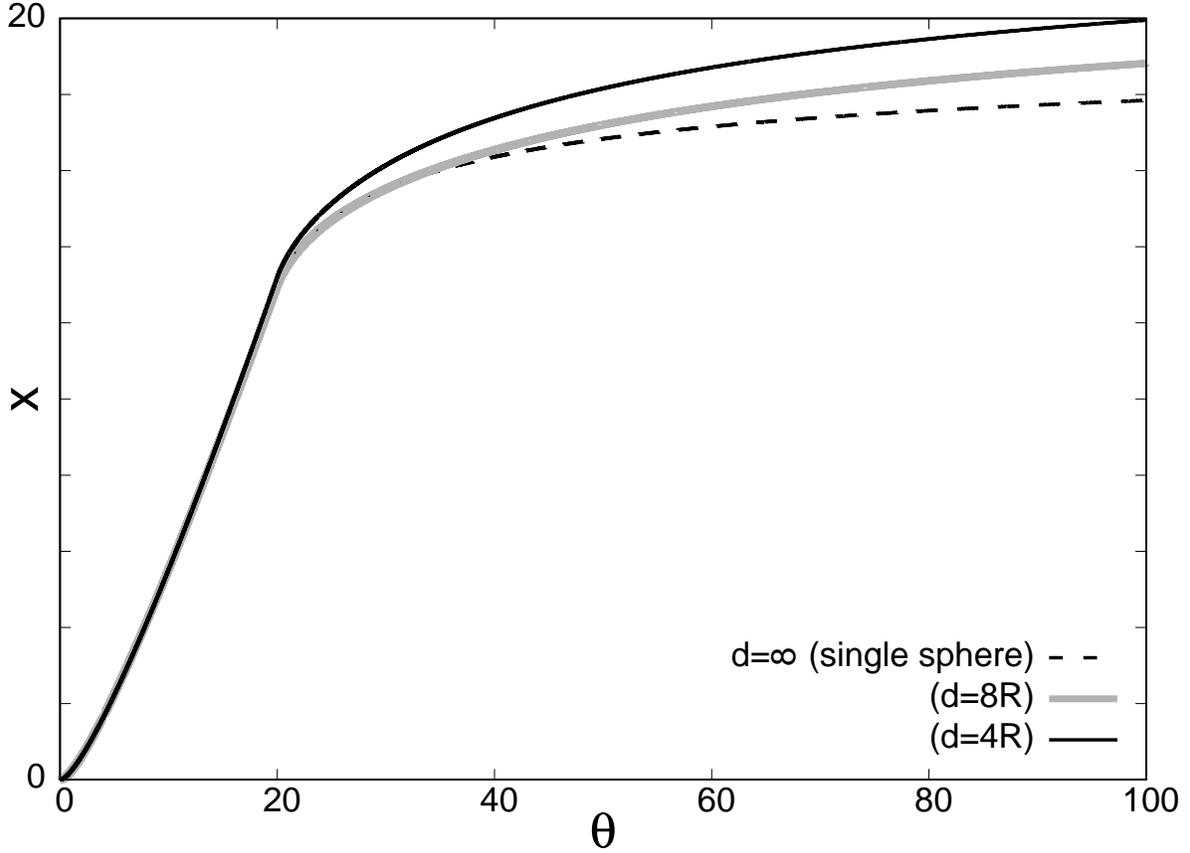}
\caption{The non-dimensional positions $x(\theta)$ of two sphere moving perpendicular to the center-to-center line, are compared for $d=\infty$ (single sphere), $d=8R$, and $d=4R$, for $T_{\rm pulse}=20 \tau_B$ and $\rho_s=\rho_f$}	
\label{per1} 
\end{figure}

\begin{figure}
\includegraphics[width=0.8\textwidth]{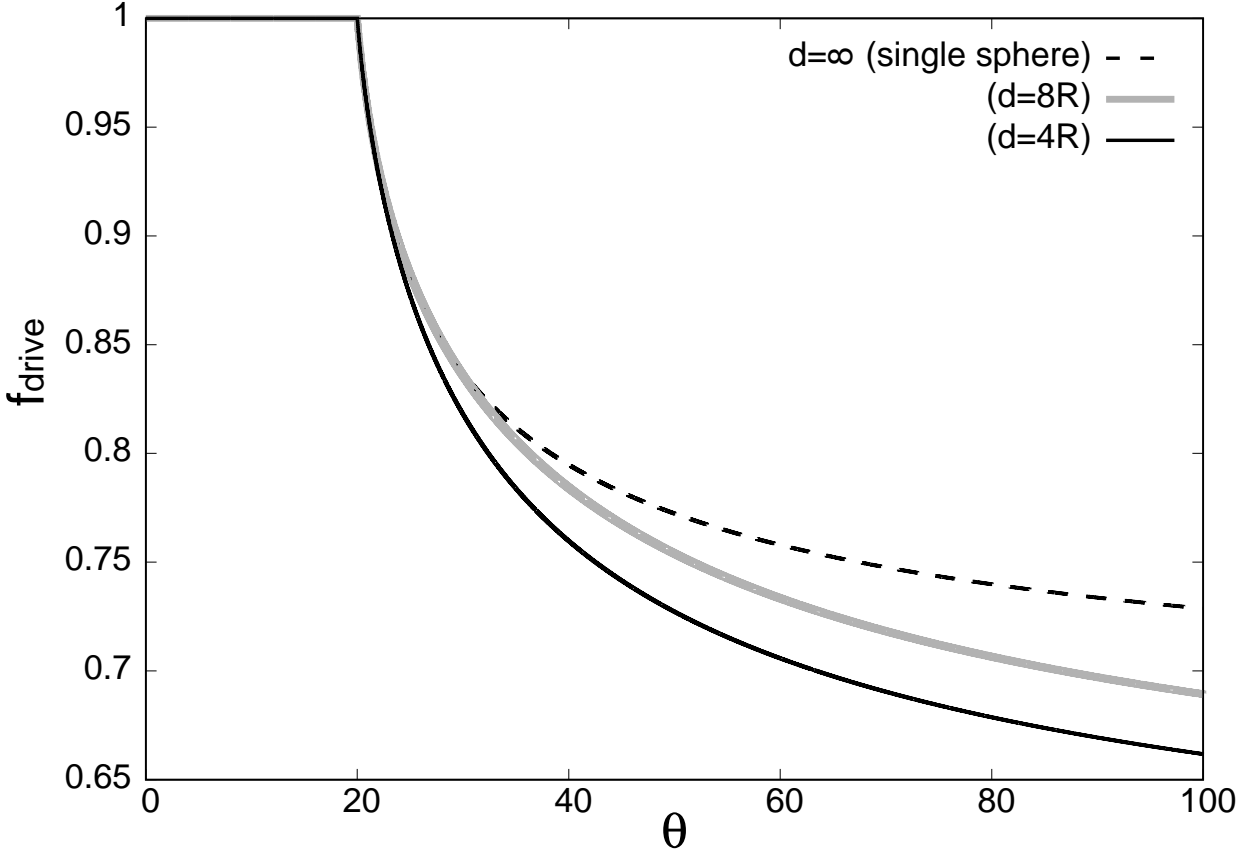}
	\caption{The effective transport drive force $f_{\rm drive}(\theta)$ of two sphere moving perpendicular to the center-to-center line, are compared for $d=\infty$ (single sphere), $d=8R$, and $d=4R$, for $T_{\rm pulse}=20 \tau_B$ and $\rho_s=\rho_f$}	
\label{per2} 
\end{figure}

 In the main text, we focused on the case where the spheres move along the line connecting their centers. We may also consider two identical spheres moving perpendicular to the  center-to-center axis, as shown in Fig.~\ref{sch2}.
The resulting force exerted by the fluid on each sphere is~\cite{2sp}
\begin{eqnarray}
        F_{\rm fluid} &=&  - \frac{ 6 \pi \eta R }{1+\frac{3}{4}\epsilon +\frac{1}{2}\epsilon^3} v
 -  \frac{ 2 \pi R^3 \rho_f}{3(1-\frac{3}{2} \epsilon^3)} \dot v \nonumber\\
&&- 6 \pi \eta R  \int^t_{-\infty} dt' \dot v(t')g_\epsilon(t-t'), \label{per21}
\end{eqnarray}
where  $g_\epsilon(t)$ is formally given as a Laplace transform~\cite{2sp},
\begin{equation}
g_\epsilon(t) = \frac{1}{\pi}\int_0^\infty \hat g_\epsilon(s) e^{-s t/\tau_\nu} ds \label{intp}
\end{equation}
where
\begin{equation}
\hat g_\epsilon(s)={\rm Im}\left[\frac{A(s)}{C_\epsilon(s)}\right] 
\end{equation}
with 
 \begin{eqnarray}
A(s) &=&  \left(1 -  i \sqrt{s} - \frac{s}{9}\right)^2\nonumber\\
	 C_\epsilon(s) &=& s^2 \left(\frac{1}{9}-\frac{\epsilon^3}{6}\right) + i \left(1 -   \epsilon^3 \right) s^{3/2}       - \left(1- \frac{5}{2}  \epsilon^3 \right) s + 3 i \sqrt{s} \epsilon^3  - \frac{3}{2} \epsilon^3 \nonumber\\
	 &&+\frac{3 \epsilon}{2} \exp\left(-i \sqrt{s}\left(1-\epsilon^{-1}\right)\right)\left[ 
     \frac{s^2}{90}  + \frac{i s^{3/2}}{6} - \frac{s}{2} - i \sqrt{s} + 1 \right]\left(-s-i \sqrt{s}\epsilon +\epsilon^2\right).
\end{eqnarray}
and $\tau_\nu \equiv \rho_f R^2/\eta$.
We note that for $\epsilon \to 0$, contribution from the term with the factor $\exp\left(-i \sqrt{s}\left(1-\epsilon^{-1}\right)\right)$ vanishes in the integral 
Eq.~(\ref{int1}), and we have
\begin{eqnarray}
g_\epsilon(t) &=& \frac{1}{\pi} \int_0^\infty {\rm Im}\left[\frac{(1 -  i \sqrt{s} - \frac{s}{9})^2}{ \frac{s^2}{9} + i  s^{3/2}       -  s}\right] e^{-s t/\tau_\nu} ds \nonumber\\
&=&  \frac{1}{\pi} \int_0^\infty \frac{1}{ \sqrt{s}} e^{-s t/\tau_\nu} ds = \sqrt{\frac{\tau_\nu}{\pi t}},
\end{eqnarray}
recovering the memory kernel for one sphere~\cite{bous,basset,oseen}.   
The qualitative behavior for two  neighboring spheres moving in the direction perpendicular to the central line is similar to those moving along the central line, in that the transport is enhanced, as shown in Figs.~\ref{per1} and \ref{per2}, where bth positions and the transport drive force are compared. We again find about 13 \% reduction in transport drive force for $d=4R$.   

\section{Solution of integro-differential equation}\label{intdf}
We numerically solved the integro-differential Eq.~(\ref{bbo3}), rewritten as
\begin{equation}
  a(\theta) = - F(\epsilon) G(\epsilon) u(\theta) - F(\epsilon) \int^\theta_{0} d \theta' a(\theta') \tilde h_\epsilon(\theta-\theta') + F(\epsilon) f(\theta). \label{bbo4}
\end{equation}
where we assumed that $v=0$ for $\theta \leq 0$,
with
\begin{eqnarray}
a(\theta) &\equiv& \frac{d u}{d \theta} \nonumber\\
\tilde h_\epsilon(\theta) &\equiv& h_\epsilon(\tau_B \theta) \nonumber\\
F(\epsilon)&\equiv& \frac{(1+3 \epsilon^3)(2  \rho_s +  \rho_f)}{2  \rho_s (1+3 \epsilon^3)+ \rho_f} \nonumber\\
 G(\epsilon) &\equiv&  \frac{1}{1+\frac{3}{2}\epsilon - \epsilon^3}.
\end{eqnarray}
The integral in Eq.~(\ref{bbo4}) is performed by using the trapezoidal rule, where we compute an integral of the form $\int_a^b f(x) dx$ by first discretizing the interval $[a,b]$ into $N$ subintervals, and approximating the integral in each subinterval as an area of the trapezoid,
\begin{equation}
\int_{a+k \Delta x}^{a + (k+1)\Delta x}  f(x) dx \simeq \Delta x\left[\frac{1}{2} f(a+k \Delta x) + \frac{1}{2} f(a+(k + 1) \Delta x\right]\label{tr1}
\end{equation}
where $\Delta x \equiv (b-a)/N$.
Care must be taken when the function $f(x)$ is divergent at a boundary, say at $x=b$. Here, we cannot use Eq.~(\ref{tr1}) at the subinterval $[b-\Delta x,b]$, so the corresponding integral must be treated separately. We utilize the asymptotic form for the function $f(x)$ for $x \to b$,
\begin{equation}
	f(x) \sim \frac{A}{(b-x)^\alpha}\quad (0 < \alpha < 1)
\end{equation}
to make the approximation
\begin{equation}
	\int_{b-\Delta x}^b f(x) dx \simeq \int_{b-\Delta x}^b  \frac{A}{(b-x)^\alpha} = \frac{A (b-\Delta x)^{1-\alpha}}{1-\alpha}. \label{asy}
\end{equation}
By using Eqs.~(\ref{tr1}) and (\ref{asy}), we obtain
\begin{equation}
	\int_a^b f(x) dx \simeq \Delta x \left[\frac{1}{2} f(a) + \sum_{k=1}^{N-2} f(a+k \Delta x) + \frac{1}{2} f(b-\Delta x)\right] + \frac{A (b-\Delta x)^{1-\alpha}}{1-\alpha}. \label{tr2}
\end{equation}

 For the integral in Eq.~(\ref{bbo4}), we have
\begin{eqnarray}
\int^\theta_{0} d \theta' a(\theta') \tilde h_\epsilon(\theta-\theta') &\simeq&  \Delta \theta \left[\frac{1}{2} a(0)\tilde h_\epsilon(\theta) + \sum_{k=1}^{N-2} a( k \Delta \theta) \tilde  h_\epsilon(\theta - k \Delta \theta)  + \frac{1}{2} a( \theta - \Delta \theta)  \tilde h_\epsilon(\Delta \theta)  \right] \nonumber\\
	&&+ \int_{\theta-\Delta \theta}^\theta  a(\theta') \tilde h_\epsilon(\theta-\theta') d \theta'. \label{midapp}
\end{eqnarray}
The last integral must be treated separately using Eq.~(\ref{asy}) since $\tilde h_\epsilon(\theta)$ diverges at $\theta=0$. Therefore we have to find the asymptotic form $\tilde h_\epsilon(\theta)$ for $\theta \to 0$. In fact, from Eq.~(\ref{int1}), we see that $h_\epsilon(t)$ is dominated by the value of $\hat h_\epsilon(s)$ at $s\to \infty$ as $t \to 0$. The oscillatory contribution containing the factor $e^{-i \sqrt{s}(1-\epsilon^{-1})}$ vanishes as $s \to \infty$, and we have
\begin{eqnarray}
	h_\epsilon(t) &\simeq& \frac{1}{\pi}\int_0^\infty {\rm Im}\left[\frac{(1 -  i \sqrt{s} - \frac{s}{9})^2}{s^2 (\frac{1}{9}+\frac{\epsilon^3}{3}) + i (1 + 2 \epsilon^3) s^{3/2}      + O(s) }\right] e^{-s t/\tau_\nu} ds \nonumber\\
&=& \frac{1}{\pi}\int_0^\infty {\rm Im}\left[\frac{s^2 + 18 i s^{3/2} + O(s) }{s^2 (9+27\epsilon^3) + i 81 (1 + 2 \epsilon^3) s^{3/2}      + O(s) }\right] e^{-s t/\tau_\nu} ds\nonumber\\
&=& \frac{1}{\pi}\int_0^\infty {\rm Im}\left[\frac{\left(s^2 + 18 i s^{3/2} + O(s)\right)\left(s^2 (9+27\epsilon^3) - i 81 (1 + 2 \epsilon^3) s^{3/2}      + O(s) \right) }{s^4 (9+27\epsilon^3)^2   + O(s^3) }\right] e^{-s t/\tau_\nu} ds\nonumber\\
&=& \frac{1}{\pi}\int_0^\infty {\rm Im}\left[\frac{ (1/9+\epsilon^3/3) +  i (1 + 4 \epsilon^3) s^{-1/2}      + O(s^{-1})  }{(1+3\epsilon^3)^2   + O(s^{-{1/2}}) }\right] e^{-s t/\tau_\nu} ds \nonumber\\
&\simeq& \frac{1}{\pi}\int_0^\infty \frac{1 + 4 \epsilon^3}{(1+3\epsilon^3)^2 \sqrt{s}} e^{-s t/\tau_\nu} ds =  \frac{1 + 4 \epsilon^3}{(1+3\epsilon^3)^2}\sqrt{\frac{\tau_\nu}{\pi t}},
\end{eqnarray}
and consequently
\begin{eqnarray}
\int_{\theta-\Delta \theta}^\theta   a(\theta') \tilde h_\epsilon(\theta-\theta') d \theta' &\simeq&  \int_{\theta-\Delta \theta}^\theta   \frac{(1 + 4 \epsilon^3) a(\theta')}{(1+3\epsilon^3)^2} \sqrt{\frac{\tau_\nu}{\pi \tau_B (\theta-\theta')}} d \theta'
\nonumber\\
&\simeq&  a(\theta) \int_{\theta-\Delta \theta}^\theta   \frac{(1 + 4 \epsilon^3) }{(1+3\epsilon^3)^2} \sqrt{\frac{\tau_\nu}{\pi \tau_B (\theta-\theta')}} d \theta' \nonumber\\
	&=&  \frac{2(1 + 4 \epsilon^3) }{(1+3\epsilon^3)^2} \sqrt{\frac{\tau_\nu  \Delta \theta}{\pi \tau_B}} a(\theta). \label{tailapp}
\end{eqnarray}
Note that the right-hand side of Eq.~(\ref{bbo4}) contains $u(\theta)$ that is undetermined at the time when computing $a(\theta)$. It is to be computed using the trapezoidal rule  
\begin{equation}
	u(\theta)=u(\theta-\Delta \theta) + \frac{\Delta \theta}{2} \left[a(\theta)+a(\theta-\Delta \theta)\right]. \label{trenew}
\end{equation}
We simply substitute Eq.~(\ref{trenew}) into Eq.~(\ref{bbo4}), along with and Eqs.~(\ref{midapp}) and (\ref{tailapp}) to get 
\begin{eqnarray}
	a(\theta) &=& - F(\epsilon) G(\epsilon) u(\theta-\Delta \theta) - \frac{\Delta \theta}{2} F(\epsilon) G(\epsilon) \left[a(\theta) + a(\theta-\Delta \theta)\right] \nonumber\\
&&- F(\epsilon)  \Delta \theta \left[\frac{1}{2} a(0)\tilde h_\epsilon(\theta) + \sum_{k=1}^{N-2} a( k \Delta \theta) \tilde h_\epsilon(\theta - k \Delta \theta)  + \frac{1}{2} a( \theta - \Delta \theta) \tilde h_\epsilon(\Delta \theta)  \right] \nonumber\\
	&&- F(\epsilon) \frac{2(1 + 4 \epsilon^3) }{(1+3\epsilon^3)^2}\sqrt{\frac{\tau_\nu  \Delta \theta}{\pi \tau_B}} a(\theta) \nonumber\\
&&+ F(\epsilon) f(\theta).\label{bbo5}
\end{eqnarray}
Moving the term proportional to $a(\theta)$ to the right-hand side to the left-hand side and solving for $a(\theta)$, we get
\begin{eqnarray}
	a(\theta) &=& - H(\epsilon) F(\epsilon) G(\epsilon) u(\theta-\Delta \theta)  \nonumber\\
	&&- H(\epsilon) F(\epsilon)  \Delta \theta \left[\frac{1}{2} a(0)\tilde h_\epsilon(\theta) + \sum_{k=1}^{N-2} a( k \Delta \theta) \tilde h_\epsilon(\theta - k \Delta \theta)  + \frac{1}{2} a( \theta - \Delta \theta) \left(\tilde h_\epsilon(\Delta \theta) + G(\epsilon) \right)  \right] \nonumber\\
&&+ H(\epsilon)F(\epsilon) f(\theta),\label{bbo6}
\end{eqnarray}
where
\begin{equation}
	H(\epsilon) \equiv \left[1 + F(\epsilon) \frac{2(1 + 4 \epsilon^3) }{(1+3\epsilon^3)^2}\sqrt{\frac{\tau_\nu  \Delta \theta}{\pi \tau_B}} + \frac{\Delta \theta}{2} F(\epsilon) G(\epsilon)\right]^{-1}. \label{new}
\end{equation} 
Eq.~(\ref{bbo6}) allows us to compute $a(\theta)$ in terms of $u(\theta-\Delta \theta)$ and $a(0), a(\Delta \theta), \cdots a(\theta - \Delta \theta)$, by storing $a(\theta)$'s as arrays during the computation. Once $a(\theta)$ is obtained, $u(\theta)$ can be obtained by Eq.~(\ref{trenew}).

We also need to compute the integral in Eq.~(\ref{int1}) in order to obtain $\tilde h_\epsilon(\theta)$. The upper limit of the integral is infinity, so we truncate the integral when the integrand is sufficiently small. In other words, we truncate the region with $ e^{-s \tau_B \theta/\tau_\nu} < \delta$. The remaining integral is obtained numerically by the trapezoidal rule. It would be computationally inefficient to perform the integral in Eq.~(\ref{int1}) each time we compute $a(\theta)$. Therefore, we compute $\tilde h_\epsilon(0),\tilde h_\epsilon(\Delta \theta),\cdots \tilde h_\epsilon(N \Delta \theta)$ at the start of the computation and store them as arrays, where $N\Delta \theta$ is the upper limit of $\theta$ that $a(\theta)$ will be computed. 

\begin{figure}
\includegraphics[width=0.8\textwidth]{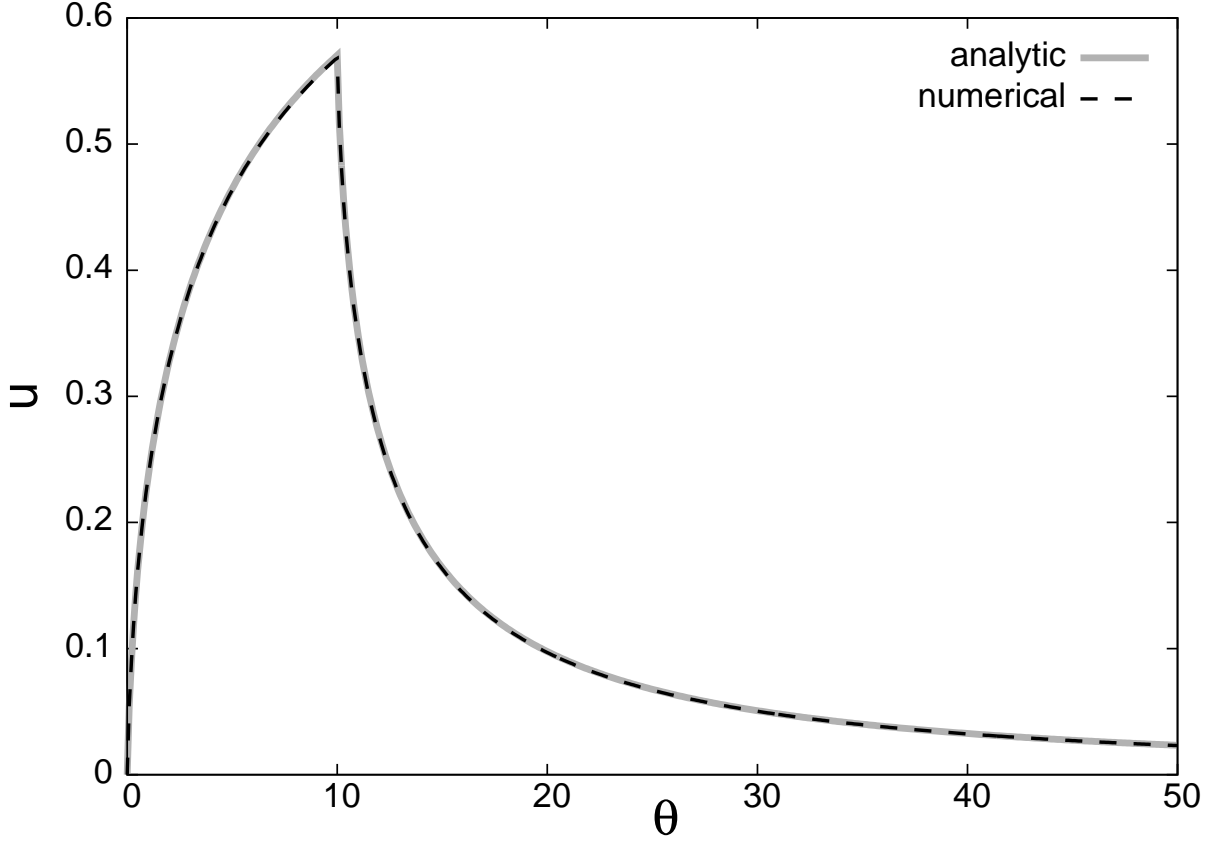}
\caption{The velocity of a single sphere as the function of time. Analytic and numerical solutions are compared for $\rho_s=0$. The integration parameters are $\Delta \theta=10^{-3}$, $\Delta s=10^{-4}$, and $\delta=10^{-4}$.}
\label{vel1} 
\end{figure}

For an isolated sphere, an analytic solution of Eq.~(\ref{bbo}) for a constant pull is known~\cite{1sp} which can be compared with our numerical solution for $\epsilon=0$ to assess the accuracy of our method. We found that using $\Delta \theta =0.001$, $\Delta s = 0.0001$, and $\delta=0.0001$ yields reasonably accurate solution, as can be seen in Fig.~\ref{vel1} where the numerical and the analytic solutions for the velocities are compared for $\rho_s=0$\footnote{The analytic solution for $\rho_s=\rho_f$ is expressed in terms of error function with complex argument~\cite{1sp}, 
For convenience, we therefore use $\rho_s=0$ for the purpose of plotting the analytic solution. }. These parameters were also used for performing the numerical integration of the two-sphere equation.

\subsection{Assessment of the accuracy of the truncation in Eq.~(\ref{bbo21})}
\begin{figure}
\includegraphics[width=0.8\textwidth]{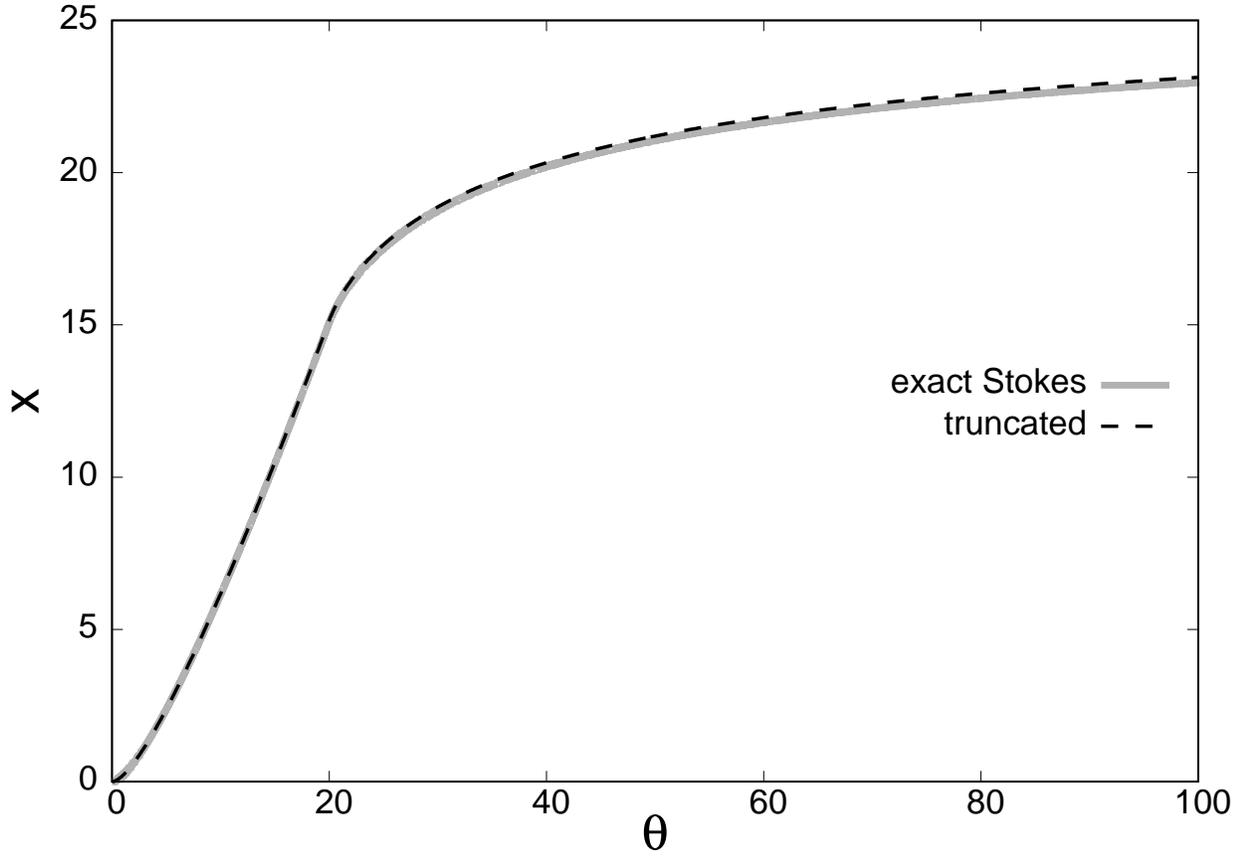}
	\caption{The position of each sphere as the function of time. Result obtained from the truncated expression and the exact one for the Stokes drag are compared, for $t=20 \tau_B$ and $\rho_s = \rho_f$.} 
\label{jeff} 
\end{figure}

As mentioned in the main text, Eq.~(\ref{bbo3}) is obtained by a truncation where the error is considered to be of order $O(\epsilon^4)$ where $\epsilon \equiv R/d$. It was argued in Ref.~\cite{2sp} that Eq.~(\ref{bbo3}) is reasonably accurate for $\epsilon \lesssim 0.25$, by comparing the Stokes drag in 
Eq.~(\ref{bbo3}), 
\begin{equation}
	f_{\rm Stokes}^{(\rm trunc)} = -\frac{u}{1+\frac{3}{2}\epsilon-\epsilon^3},
\end{equation}
with the known exact form~\cite{jf}\footnote{There is an error in the overall prefactor of the final expression in Ref. \cite{jf} where $2/3$ is given instead of $4/3$. We could check that $4/3$ is the correct one by carefully following their derivation.}
\begin{equation}
	f_{\rm Stokes}^{(\rm exact)} = -\frac{4u}{3}\sum_{n=1}^\infty \sinh \alpha \frac{n(n+1)}{(2n-1)(2n+3)} 
	\left[1-\frac{4 \sinh^2 (n+\frac{1}{2})\alpha - (2n+1)^2 \sinh^2 \alpha}{ 2 \sinh (2n+1) \alpha + (2n+1) \sinh  2 \alpha }\right],
\end{equation}
where 
\begin{equation}
	2 \cosh \alpha \equiv \epsilon^{-1}.
\end{equation}
The numerical result is quite robust as we replace $f^{\rm (trunc)}_{\rm Stokes}$ by $f^{\rm (exact)}_{\rm Stokes}$ up to $\epsilon=0.25$, as shown in Fig.~\ref{jeff} where the results for the position $x$ are compared for $t=20\tau_B$ and $\rho_s = \rho_f$. This provides circumstantial evidence that Eq.~(\ref{bbo3}) is quite accurate for $\epsilon \lesssim 0.25$. We however do not use $f_{\rm Stokes}^{\rm exact}$ as we avoid {\it ad hoc} mixing of exact and truncated expressions.
\end{document}